**$Mg_xZn_{1-x}O$ contact to $CuGa_3Se_5$ absorber for photovoltaic and photoelectrochemical devices**

Imran S. Khan[1], Christopher P. Muzzillo[1], Craig L. Perkins[1], Andrew G. Norman[1], James Young[1], Nicolas Gaillard[2], and Andriy Zakutayev[1]

[1] National Renewable Energy Laboratory, Golden, CO, USA

[2] Hawaii Natural Energy Institute, University of Hawaii, Honolulu

## Abstract:

$CuGa_3Se_5$ is a promising candidate material with wide band gap for top cells in tandem photovoltaic (PV) and photoelectrochemical (PEC) devices. However, traditional CdS contact layers used with other chalcopyrite absorbers are not suitable for $CuGa_3Se_5$ due to the higher position of its conduction band minimum. $Mg_xZn_{1-x}O$ is a transparent oxide with adjustable band gap and conduction band position as a function of magnesium composition, but its direct application is hindered by $CuGa_3Se_5$ surface oxidation. Here, $Mg_xZn_{1-x}O$ is investigated as a contact (n-type 'buffer' or 'window') material to $CuGa_3Se_5$ absorbers pretreated in $Cd^{2+}$ solution, and an onset potential close to 1 V vs RHE in 10 mM hexaammineruthenium (III) chloride electrolyte is demonstrated. The $Cd^{2+}$ surface treatment changes the chemical composition and electronic structure of the $CuGa_3Se_5$ surface, as demonstrated by photoelectron spectroscopy measurements. The performance of $CuGa_3Se_5$ absorber with $Cd^{2+}$ treated surface in the solid-state test structure depends on the Zn/Mg ratio in the $Mg_xZn_{1-x}O$ layer. The measured open circuit voltage close to 1 V is promising for tandem PEC water splitting with $CuGa_3Se_5/Mg_xZn_{1-x}O$ top cells.

# 1. Introduction:

Hydrogen is considered one of the most promising means of storing renewable energy because it is the most abundant element in the world, has the highest energy density among non-nuclear fuels, and is a zero-emission fuel [1]. Solar photoelectrochemical (PEC) water splitting can be used to produce hydrogen, but a commercially viable PEC device technology remains elusive despite long research history [2]. Tandem monolithic PEC devices with a multijunction hybrid photoelectrode [3] consisting of group III-V semiconductors have been able to reach direct water splitting solar-to-hydrogen (STH) efficiency above 19% [4]. However, large scale commercial adoption of such technology is limited by high processing costs and finite durability during PEC water splitting. To maintain high efficiency and reduce processing cost, tandem PEC devices with chalcopyrite absorber materials can be adopted from photovoltaic (PV) research. These chalcopyrites have lower deposition cost and higher defect tolerance, which together with bandgap tunability makes them attractive for application in solar water splitting. Indeed, wide bandgap chalcopyrite $CuGaSe_2$ [5] [6], $CuGa(S,Se)_2$ [7] and $Cu(In,Ga)S_2$ [8] photocathodes have shown promising PEC performance and stability, and emerging wide bandgap $CuGa_3Se_5$ has been recently proposed [9].

$CuGa_3Se_5$ is an ordered-vacancy compound derived from the chalcopyrite structure of the well-known $CuGaSe_2$ absorber material by substituting Ga at Cu sites and leaving Cu sites vacant, $[2V_{Cu}^{-1} + Ga_{Cu}^{+2}]$ [10]. Reducing the Cu/Ga composition widens the band gap by reducing the valence band energy by ~0.2 eV, because of lower contribution from the Cu 3d states to the valence bands [11]. Due to its ideal bandgap of 1.84 eV and suitable conduction band alignment for $H_2$ evolution, $CuGa_3Se_5$ is a promising absorber material candidate for top cell applications in tandem PEC water splitting devices. Calculations show that if paired with a 1.23 eV bandgap absorber in a tandem PEC cell, $CuGa_3Se_5$ could potentially lead to STH efficiency as high as 22.8% [12]. Initially a $CuGa_3Se_5$/ZnS/Pt device with current output of 8 mA/cm$^2$ (0V vs RHE, 3-electrode) was reported [13], and subsequently the photocurrent density was increased up to 9.3 mA/cm$^2$ for a mixed phase of $CuGaSe_2$ and $CuGa_3Se_5$ with CdS-modified surface and Pt catalyst [14]. More recently, 17 days of continuous water splitting operation for a bare $CuGa_3Se_5$ absorber photocathode with ~12 mA/cm$^2$ photocurrent at −1 V vs RHE have been demonstrated by our team, suggesting promising durability [9]. These encouraging outcomes clearly point out the need for further fundamental study of this absorber and its interface with other materials.

Traditional chalcopyrites $CuGaSe_2$ and ordered-vacancy $CuGa_3Se_5$ absorbers are often interfaced with CdS as a contact material (also known as n-type 'buffer' or 'window') to create the pn-junction. However, CdS contact layers suffer from short wavelength absorption and instability in electrolyte solution. In addition, CdS has a cliff-like ~0.2 eV conduction band (CB) offset with stoichiometric $CuGaSe_2$, and a similar CB offset is expected for $CuGa_3Se_5$ [15]. If the conduction band minimum (CBM) of the contact is lower than that of the absorber (a 'cliff' type offset), the device suffers from lower photovoltage, increased interface recombination, and other detrimental effect to device performance [16] [17] [18]. If a CBM of the contact is more than 0.3 eV above that of the absorber (a 'spike' type offset), the resulting barrier impedes the collection of photo-generated carriers [19] [20] [21].

To address these challenges, $Mg_xZn_{1-x}O$ (MZO) [22] [23] with tunable conduction band position as a function of Mg content has been proposed as an attractive n-type contact layer. MZO thin film was demonstrated with up to x = 0.46 grown by RF co-sputtering without any phase segregation resulting in a bandgap of up to 4.2 eV, while ZnO has a bandgap of 3.24 eV [22]. Combinatorial studies explored the composition spreads of MZO with different deposition methods, such as pulsed laser deposition [24] [25] and chemical vapor deposition [26]. In a previous combinatorial study, we showed that the conduction band position could be tuned by 0.5 eV as Mg concentration changes from 4 to 12% [27], suggesting that it might be a suitable contact to $CuGa_3Se_5$ absorber. Integration of MZO as contact material resulted in significant efficiency improvements in different solar cell device technologies such as CdTe [28] and CIGS [29] [30], as well as $CuGaSe_2$ [31] that likely has similar CB position to $CuGa_3Se_5$.

In this study, MZO was investigated as the contact layer material for $CuGa_3Se_5$ absorber-based PV and PEC devices. Structural, optical and electrical properties of MZO thin films were studied as a function of different experimental conditions such as Mg composition, Ga doping, substrate temperature and deposition ambient. MZO depositions were performed by combinatorial radio frequency (RF) sputtering. For functional $CuGa_3Se_5$/MZO PV device, absorber surface pretreatment with $Cd^{2+}$ solution was crucial because it removed surface oxidation, led to Cd incorporation, and possibly changed the surface conductivity type. MZO deposition and the surface pretreatment conditions were optimized for solid state solar cell performance. The outcome was a significant improvement in open circuit voltage (up to 920 mV) compared to conventional CdS-contact $CuGa_3Se_5$ devices (~730 mV). Replacement of CdS also improved quantum efficiency in the blue region of the spectrum. The PEC characteristics of $CuGa_3Se_5$/MZO as photocathode are tested with hexaammineruthenium (III) chloride sacrificial redox electrolyte, which exhibited an onset potential near 1 V vs reversible hydrogen electrode (RHE). These outcomes indicate that $CuGa_3Se_5/Mg_xZn_{1-x}O$ could serve as an efficient top cell for tandem PV and PEC water splitting devices.

## 2. Experimental Methods:

### 2.1 Material synthesis and measurements

Ga-doped $Mg_xZn_{1-x}O$ thin film sample libraries with orthogonal composition gradients of Mg and Ga were deposited by combinatorial RF magnetron sputtering from ZnO, Mg and $Ga_2O_3$ targets (Figure 1a). 50x50 mm Eagle XG glass substrates were cleaned with laboratory grade detergent followed by sonication in warm DI water, acetone and isopropanol. MZO depositions were done at a pressure of 3 mTorr in $Ar/O_2$ atmosphere where the total gas flow rate was fixed at 16 sccm. The chamber base pressure was $5 \times 10^{-7}$ Torr. $O_2$ flow, found to be crucial for good quality transparent films, was varied from 0.5 to 2% of the total gas flow for different depositions. A premixed 5% $O_2$ in Ar cylinder was used as $O_2$ source for precise flow control. A gas ring around the substrate carrying the $O_2$ lines ensured uniform $O_2$ containing environment across the substrate. The samples were mounted on a temperature calibrated Inconel substrate holder and heated by a radiative heater. The substrate temperature was varied from room temperature to 200 °C. The depositions were performed for 120 min, that resulted in film thicknesses in the 80 to 120 nm range.

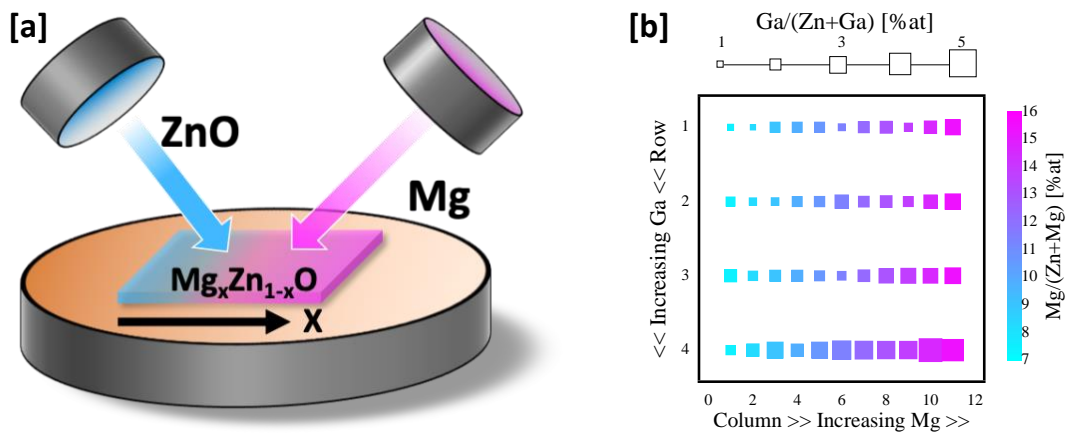

Figure 1. [a] Combinatorial sputter deposition of $Mg_xZn_{1-x}O$ thin films. [b] 4x11 composition measurement grid for $Ga:Mg_xZn_{1-x}O$ films on a 50mm × 50mm glass substrate.

Each sample was characterized at 4x11 grid points with the following spatially resolved methods (Figure 1b). X-ray diffraction (XRD) patterns for the MZO thin films were obtained using a Bruker D8 Discover XRD instrument. Electrical resistivity/conductivity data was measured by a custom four-point probe system; the highest measurable sheet resistance by the equipment is ~$5x10^7$ Ω/□. Optical absorption was measured with a custom UV/Vis/NIR Spectroscopy system equipped with an OceanOptics DH-2000-BAL deuterium-halogen light source and a StellerNet EPP2000-UVN-SR spectrometer. The chemical compositions of the films, atomic Ga/(Ga+Zn) ratio, were measured with a Fisherscope XUV-733 X-ray fluorescence (XRF) instrument. Mg, being a light element, was quantified using Rutherford backscattering spectroscopy (RBS) and energy dispersive x-ray spectrometry (EDX). EDX data was taken using an acceleration voltage of 5.0 keV. For Mg quantification purpose, 10x10 mm glassy carbon or silicon witness substrates were placed at the 2[nd] row of the 4x11 sample library grid. The use of such impurity free witness substrates reduced background noise in quantification of light elements in small amounts. Experimental combinatorial data collected in this study were managed, analyzed and displayed using our publicly available COMBIgor software package for Igor Pro [32], and will be made available through the High Throughput Experimental Materials Database (HTEM DB) [33].

Kelvin probe measurement system from KP technology was used to determine the work function for both the MZO (on doped Si substrates) and $CuGa_3Se_5$ (on Mo coated glass substrates) films. A gold reference (work function 5.1 eV) in air was used to quantify the absolute value of the surface potentials. X-ray photoelectron spectroscopy (XPS) was performed on the $CuGa_3Se_5$ films using monochromatic Al Kα radiation and a pass energy of 29 eV. The spectrometer binding energy scale was calibrated at high and low energy using clean gold and copper foils and known transition energies. Data analysis and peak fitting were performed using a combination of Igor and PHI MultiPak.

Cross-section scanning electron microscopy (SEM) image of the device was taken with a Hitachi S-4800 SEM instrument operated at 2kV. Cross-section transmission electron microscopy (TEM) specimens were prepared using the focused ion beam (FIB) lift out technique with the final $Ga^+$ ion milling performed at 3 kV. $Ga^+$ ion FIB damage was subsequently removed using low energy (< 1kV) $Ar^+$ ion milling in a Fischione Nanomill with the sample cooled by liquid nitrogen. Scanning transmission electron microscopy (STEM) imaging and EDX mapping analysis were performed in a FEI Tecnai F20 UltraTwin field emitting gun STEM operated at 200 kV and equipped with an EDAX Octane T Optima Si drift detector (SDD) EDX system.

**2.2 Device fabrication and characterization**

Substrates for device fabrication were soda-lime glass. Mo back contact was deposited by Direct Current sputtering. Near stoichiometric $CuGa_3Se_5$ thin films with Cu/Ga = 0.36 composition were deposited by three-stage co-evaporation (Ga-Se in the 1$^{st}$ stage, Cu-Se in the 2$^{nd}$ stage, and Ga-Se in the 3$^{rd}$ stage) at 600 °C. For comparison, a baseline device was fabricated in a similar fashion to narrow bandgap (~1.1 eV) CIGS absorber devices, more details on the process steps can be found elsewhere [9] [34]. The solution for $Cd^{2+}$ surface treatment contained 2mM $CdSO_4$ in $NH_4OH$ and DI water, and no S precursor. Only device data containing undoped MZO is presented here, as they produced superior devices.

PEC characteristics of the devices were measured by performing chopped-light linear sweep voltammograms (LSV) in a three-electrode configuration with $CuGa_3Se_5$ photocathode, Pt counter electrode, and Ag/AgCl reference electrode with 3M NaCl filling solution.

Photoelectrodes were made by indium bonding an insulated Cu wire to an exposed part of the Glass/Mo substrate. A number of different electrolyte solutions was tested, including (1M $Na_2SO_4$ + pH7 buffer), (NaOH + $H_2SO_4$ + 0.5M $Na_2CO_3$, pH = 9.6), and (0.5M $Na_2SO_4$ + 0.25M $KH_2PO_4$ + 0.25M $K_2HPO_4$, pH = 6.7), yet the $CuGa_3Se_5$/MZO thin films were found to be unstable in all of these solutions with the photocurrent limited due to charge transport. As such, LSV analyses were performed in a solution containing 10 mM hexaammineruthenium (III) chloride, 0.5M KCl and pH7 buffer. During the measurements, the samples were illuminated with a 300 W Xenon arc lamp (Newport) through an AM 1.5G filter (Oriel), simulating one-sun equivalent illumination as adjusted using a calibrated 1.8 eV $GaInP_2$ PV reference cell. The electrode device area was defined by nonconductive epoxy (Loctite 9462), which isolated the wire and Mo substrate from the electrolyte. The device areas were measured by counting the number of pixels from digitally scanned images of the electrodes. Potentials are reported against the reversible hydrogen electrode by using the relationship $E(RHE) = E_{Ag/AgCl} + 0.059 \times pH + E_{o, Ag/AgCl}$, where $E_{o, Ag/AgCl} = 0.209V$. The standard reduction potential ($E^0$) for the hexaammineruthenium redox couple is 0.1 V vs RHE [35].

For PV device fabrication, Al:ZnO (120 nm) as transparent conductive oxide was deposited by RF magnetron sputtering, and Ni(50 nm)/Al(3μm) front grids by e-beam evaporation [34]. Solar cell performance was characterized based on photocurrent density-voltage (JV) and quantum efficiency (QE) performances. JV data for the combinatorial device library was collected with a custom made automated XY probe station under AM1.5 light from an Oriel 91194-1000 solar simulator with the samples maintained at 20 °C. External QE for select devices were measured with a Newport Oriel IQE-200.

# 3. Results and Discussions:

## 3.1 $Mg_xZn_{1-x}O$ thin film characterization:

For $CuGa_3Se_5$ device integration, combinatorial RF sputtered $Mg_xZn_{1-x}O$ was deposited with x values in the 0 to 0.15 range, and Ga/(Ga+Zn) atomic ratio from 0% to 15%. Figure 1b shows the composition profile of one such Ga:$Mg_xZn_{1-x}O$ library, and RBS data for a selected sample is presented in Figure S1. XRD patterns of the MZO films were all (0002) oriented wurtzite ZnO; no peaks related to MgO or $Ga_2O_3$ secondary phases were observed for the investigated span of Mg and Ga compositions (Figure 2). With increasing Mg composition, the ZnO (0002) peak broadened, intensity reduced and shifted towards lower angles. This shift was more significant at higher Ga/(Ga+Zn) samples. Mg atomic size is much lower than Zn or Ga. Mg incorporation at large degree could induce strain in the ZnO crystal, indicated by the peak broadening at high Mg samples.

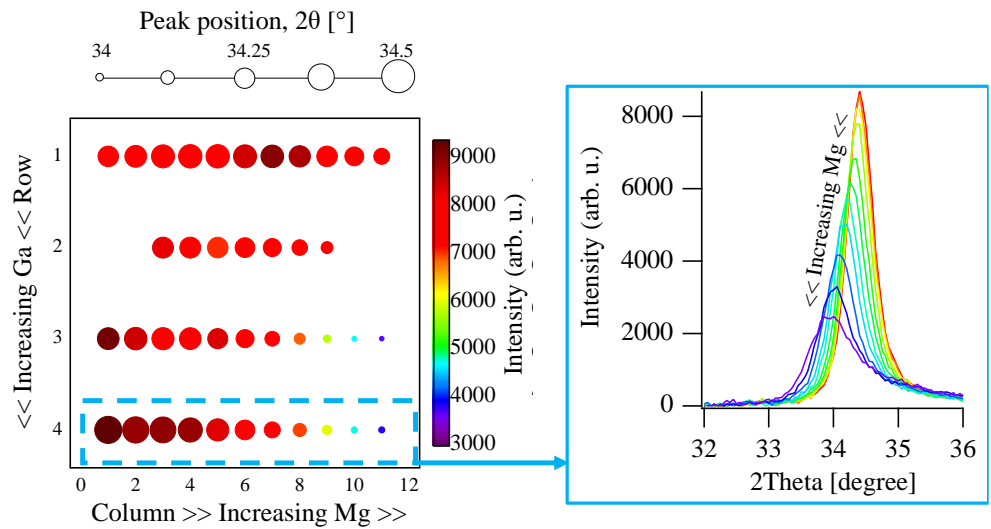

Figure 2. ZnO (0002) peak position and intensity for a Ga:$Mg_xZn_{1-x}O$ sample library deposited at 200°C.

The film conductivity for the undoped MZO films deposited at 100°C or below were too low to be measured. The conductivity for the doped MZO films were dependent on Ga and Mg compositions (Figure 3a). Conductivity decreased with increasing Mg composition, possibly due to wider bandgap and increased carrier scattering. Introduction of Ga as dopant increased conductivity, and then decreased it at higher Ga concentrations when Ga/(Ga+Zn) atomic ratio exceeded 5%. This was likely due to the reduced crystallinity of those samples, as evident from the XRD data. The highest conductivity was 20 S/cm and was observed at Ga/(Ga+Zn) value of 4%. Mg incorporation also resulted in the expected bandgap widening. Increasing Mg composition increased the optical bandgap of MZO, calculated from the Tauc plot of the absorption data from UV-Vis spectroscopy (Figure 3b). For the highest experimented Mg composition of 13%, optical band gap values up to 3.57 eV was estimated.

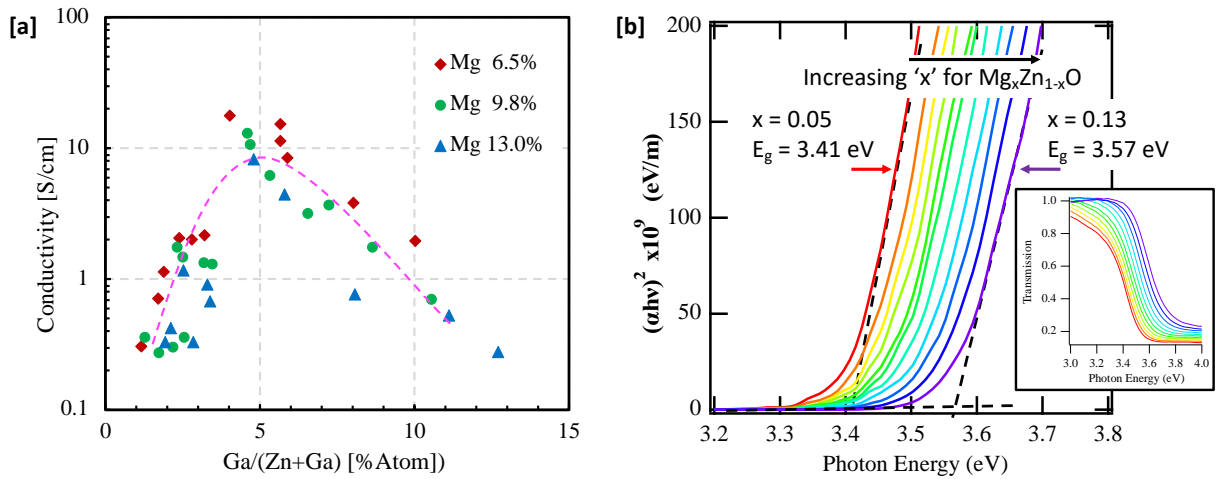

Figure 3. [a] Conductivity vs. Ga/(Ga+Zn) atomic ratio for $Mg_xZn_{1-x}O$:Ga thin films deposited at 200°C. [b] Tauc plot from the transmittance data showing the shift in the estimated optical bandgap for undoped MZO films with increasing Mg composition.

**3.2 Photoelectrochemical Characteristics:**

The PEC characteristics of the $CuGa_3Se_5$ photocathodes were investigated with chopped light linear sweep voltammetry (LSV) with a 3-electrode system under simulated AM1.5G illumination. The LSV data is shown in supplement Figure S2, for analysis performed in hexaammineruthenium (III) chloride redox mediator and KCl supporting electrolyte with pH7 buffer. It should be noted that LSV measurement performed with a sacrificial redox agent is not water splitting, however it gives an estimate of the quality of the device characteristics. There was sign of degradation during consecutive testing, however that could be due to scanning near high anodic potentials. The dark current was nearly zero. The photocurrent saturated at more cathodic potentials. The initial current transient could be due to charge transport limitation in the electrolyte and/or at the electrode surface. The photocurrent onset potentials extrapolated from the LSV curves are shown in Table 1.

Table 1. Photocurrent onset potentials from the LSV data

| Electrode | $V_{onset}$ (V vs. RHE) |
|---|---|
| $CuGa_3Se_5$ (As Deposited) | $0.57 \pm 0.02$ |
| $CuGa_3Se_5/(Cd^{2+})$ | $0.64 \pm 0.02$ |
| $CuGa_3Se_5/CdS/i\text{-}ZnO$ | $1.18 \pm 0.04$ |
| $CuGa_3Se_5/(Cd^{2+})/i\text{-}MZO$ | $1.08 \pm 0.03$ |

The untreated $Mo/CuGa_3Se_5$ electrode had a photocurrent onset potential ($V_{onset}$) of ~0.57 V vs. RHE. The Fermi level of the p-type $CuGa_3Se_5$ semiconductor would be located much lower than the solution potential (reduction potential of the redox). This would create a charge depletion layer in the semiconductor due to transfer of valence band hole to the solution. In equilibrium, this charge separation would form a built-in electric field, causing a downward band bending. The potential $V_{onset}$ is indicative of this built-in potential. An increase of $V_{onset}$ up to 0.64 V vs. RHE, for $Cd^{2+}$ solution treated $CuGa_3Se_5$ electrode indicated an increase of this band bending.

For $CuGa_3Se_5$/n-CdS/i-ZnO device, $V_{onset}$ close to 1.2 V vs. RHE was observed. This was indicative of the good interface passivation and/or extended depletion region due to increased band bending at the pn-junction formed between n-CdS and p-$CuGa_3Se_5$. The optimized $CuGa_3Se_5$/($Cd^{2+}$)/MZO devices also exhibited $V_{onset}$ potential near 1.0 V vs. RHE. However, this value for devices with $Cd^{2+}$ treated absorber surfaces was lower compared to baseline CdS devices. This difference in the onset potential could not be due to substitution of MZO for i-ZnO alone, since $CuGa_3Se_5$/MZO solid-state stacks resulted in no photo-response. Even with the absence of an evident active n-type layer, the $CuGa_3Se_5$/($Cd^{2+}$)/MZO devices still exhibited $V_{onset}$ potential above 1V vs. RHE. This encouraging result indicated that $Cd^{2+}$ treatment itself introduced a surface passivation and/or a band bending at the absorber interface. However, for ($Cd^{2+}$)-MZO (undoped) modified electrode the band bending must be lower compared to the CdS modified electrode, evident form the onset potential values. A schematic of the equilibrium band diagram for a $CuGa_3Se_5$/MZO photoelectrode is shown in the supplement Figure S2 inset.

### 3.3 Surface and Interface:

To better understand the PEC characteristics of the electrodes, the absorber surface and interface modification due to the $Cd^{2+}$ solution treatment was further investigated. Surface sensitive XPS/UPS of the as-deposited and treated $CuGa_3Se_5$ revealed how the wet treatment affected the absorber surface (Figure 4). The XPS survey spectrum of the as-deposited $CuGa_3Se_5$ is shown in the supplement Figure S3. The surface doping of as grown $CuGa_3Se_5$ appeared stable over a period of months, and was only slightly p-type (the XPS measurement on the as grown sample was repeated after 3 months, data not shown). The valence band position ($E_F$-$E_{VBM}$) value measured with monochromatic Al $K_\alpha$ and He I excitation was 0.55 eV and 0.72 eV respectively (supplement Figure S4). UPS results with He I light (21.2 eV) had a probe depth of less than 1 nm in the sample surface, while VBM measurements with Al $K_\alpha$ (1487 eV) measured up to 10 nm into the bulk of the film due to higher photoelectron kinetic energy. This indicated the presence of a downward band bending at the as-grown surface.

The elemental compositions of the $CuGa_3Se_5$ surface with different treatments is shown in Table 2. There were a number of changes that happened with both $NH_4OH$ and $Cd^{2+}$ (65°C 15 min) treatments (Figure 4). A high degree of surface oxidation is observed for as deposited absorber, along with Na diffused from the soda lime glass substrate. With both $NH_4OH$ and $Cd^{2+}$ treatments, surface Na was removed and the O quantity was reduced. The gallium $2p_{3/2}$ peak narrowed significantly with $Cd^{2+}$ treatment, probably due to the removal of gallium oxides. It is also interesting to note that the cadmium solution treatment (and likely the subsequent air exposure), caused the appearance of oxidized selenium, $Se^{4+}$. For CdTe devices a very thin layer of oxidized tellurium was found to be critical for well-passivated interfaces [36]. Such Se oxidation could also be beneficial for surface passivation of the $CuGa_3Se_5$ absorbers studied here.

No Cd could be detected on the $Cd^{2+}$ treated $CuGa_3Se_5$ films using XRF and SEM/EDX analysis techniques, which are more sensitive to the bulk than XPS. Such comparison of XPS and EDX (probes deeper, up to 100 nm) and XRF (probes up to 1000 of nm) compositions for as-grown films indicated a Cu deficiency on the surface which is likely due to the sites being replaced by Na atoms. $Cd^{2+}$ treatment introduced Cd on the absorber surface (supplement Figure S3). Exposure of the surface to aqueous cadmium sulfate likely caused an ion exchange process between cadmium and copper. Removal of Na atoms by $NH_4OH$ could allow Cd atoms to occupy these sites and change surface doping. The ($E_F$-$E_{VBM}$) values from XPS and work function values from Kelvin probe measurement for differently treated films also suggested a change in surface doping.

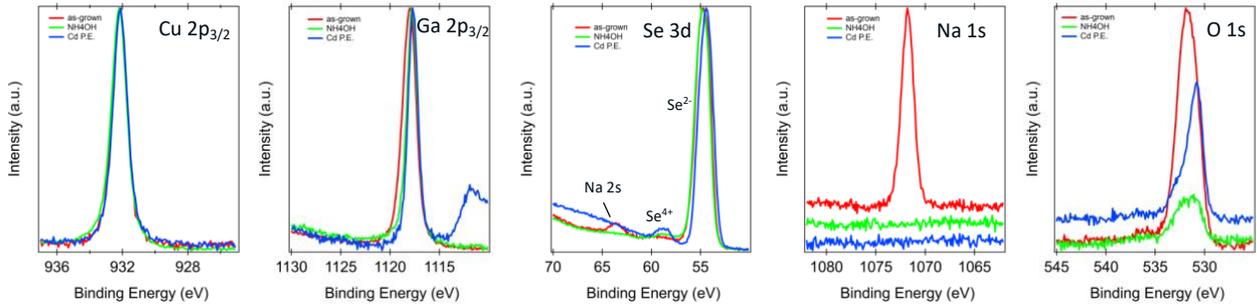

Figure 4. Elemental core levels of the $CuGa_3Se_5$ surface with different surface treatments, measured by XPS.

Table 2. XPS and EDX composition of untreated and treated $CuGa_3Se_5$ absorber.

| $CuGa_3Se_5$ | Technique | O (%) | Na (%) | Cu (%) | Ga (%) | Se (%) | Cd (%) |
|---|---|---|---|---|---|---|---|
| **As Deposited** | XPS | 41.8 | 7.9 | 3.4 | 24.5 | 22.5 | - |
| | (normalized) | | | **6.7** | 48.6 | 44.7 | - |
| | XRF | - | - | **11.2** | 32.5 | 55.6 | |
| | EDS | - | - | 14.2 | 45.1 | 40.7 | - |
| **NH$_4$OH treatment** | XPS | 10.9 | - | 18.4 | 25.1 | 45.7 | - |
| | (normalized) | | | **20.6** | 28.1 | 51.3 | - |
| **Cd$^{2+}$ treatment** | XPS | 21.2 | - | 4.8 | 19.8 | 35.8 | 18.4 |
| | (normalized) | | | **8.0** | 32.7 | 59.3 | |

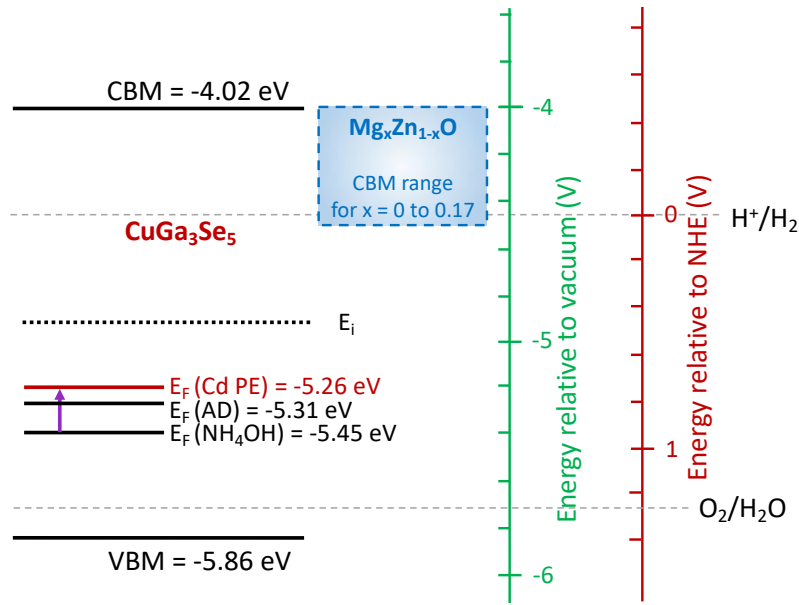

Figure 5. Energy band positions for CuGa$_3$Se$_5$ absorber with different surface treatments, in comparison with the CBM position in Mg$_x$Zn$_{1-x}$O.

Figure 5 shows the band diagram for CuGa$_3$Se$_5$ with Fermi energy level ($E_F$) positions for different treatments. The $E_F$ values with respect to valence band maximum ($E_{VBM}$) are derived from XPS measurements. The change in $E_F$ with different surface treatments were further validated with relative work function values from Kelvin probe measurements. CBM energy level was calculated using the bandgap value of 1.84 eV for CuGa$_3$Se$_5$. For as deposited bulk thin films, $E_F - E_{VBM} = 0.28$ eV was calculated using the carrier concentration of ~$2\times10^{14}$ cm$^{-3}$ from Hall effect measurement and the carrier effective masses were obtained from literature [37]. The attainable CBM range for MZO is from our previous combinatorial study [27].

An $E_F$-$E_{VBM}$ of 0.55 eV for the as deposited absorber surface suggested a downward band bending with reduced p-type conductivity. A NH$_4$OH treatment lowered the value to 0.41 eV due to the removal of surface states. $E_F$-$E_{VBM}$ increased to 0.6 eV with Cd$^{2+}$ treatment, likely due to Cd replacing Cu, creating compensating defect(s) on the absorber surface. This implied a shift in surface conductivity type towards becoming intrinsic. Cadmium did not appear to dope the surface fully n-type, unlike what was reported for CIGSe or CISe [38] [39]. CIGSe and CISe normally are p-type at the surface and after cation exchange with cadmium solution, become n-type. However, $Cd^{2+}$ treatment for $CuGa_3Se_5$ only moved the Fermi level upward, closer to being intrinsic.

### 3.4 Photovoltaic characterization:

To understand the influence of the contact CB position on the absorber performance in PEC environment, $Mg_xZn_{1-x}O$ with varying composition was integrated with $CuGa_3Se_5$ absorbers into solid state test structures with top TCO and metal contacts. For the baseline PV device ($CuGa_3Se_5$/CdS/i-ZnO material stack), the open circuit voltage ($V_{OC}$) = 730 mV, short-circuit current ($J_{SC}$) = 5.7 mA/cm$^2$, fill factor (FF) = 61%, and efficiency 2.6% were achieved. Initial devices with CdS contact layer, and MZO layer replacing i-ZnO layer showed promising results: $V_{OC}$ = 755 mV, $J_{SC}$ = 7.6 mA/cm$^2$, FF = 38.4%, and efficiency 2.1% (Supplementary Figure S5). A clear CdS layer was observed in SEM (Figure 6a), STEM/HAADF (Figure 6c) and STEM/EDX (Figure 6e) images.

$Mg_xZn_{1-x}O$ was then studied as a replacement for the CdS/i-ZnO contact stack. PV devices without any surface treatment of the as-deposited $CuGa_3Se_5$ didn't exhibit any quantifiable current generation. Solution treatments of $CuGa_3Se_5$ indicated a pathway to replace CdS with MZO, with MZO grown directly on $NH_4OH$-treated and $Cd^{2+}$ solution treated $CuGa_3Se_5$ surfaces. The $NH_4OH$ treatment of the absorber prior to MZO deposition resulted in some photovoltaic activity, although the device performance values were quite low. $Cd^{2+}$ solution treated $CuGa_3Se_5$ surfaces led to improved PV performance, and some incorporation of Cd into $CuGa_3Se_5$ surfaces.

The material stack for the PV devices with Cd-treated $CuGa_3Se_5$ absorbers and MZO contacts are shown in Figure 6b. The effect of the $Cd^{2+}$ solution treatment could not be resolved from the SEM (Figure 6b) or STEM/HAADF (Figure 6d) images, indicating that very small amount of Cd was substituted at the surface (as indicated by XPS). However, STEM/EDX elemental map of the device revealed Cd present at the $CuGa_3Se_5$/MZO interface (Figure 6f). Elemental line profiles from STEM/EDX showing the change in elemental composition across the cross section for these devices is shown in supplementary Figure S6. A small amount of Sulphur (S) is also present at this interface, which is likely due to $CdSO_4$ that was the source of $Cd^{2+}$.

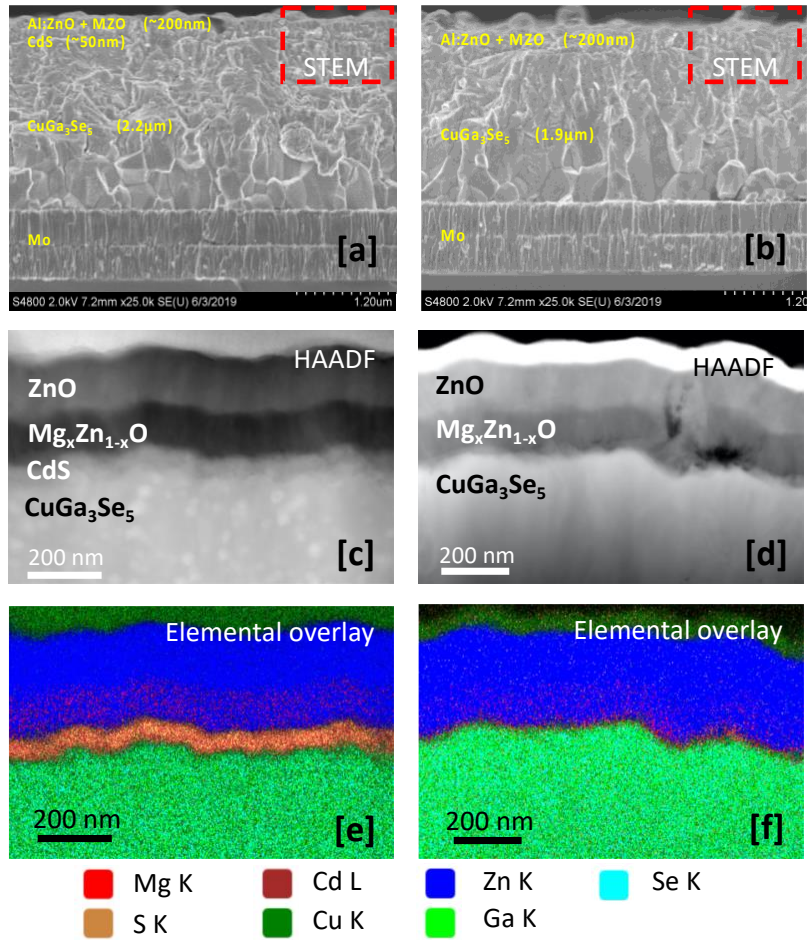

Figure 6. (a, b) SEM images with color overlay for the PV devices. (c, d) STEM HAADF images and (e, f) STEM EDX elemental maps at the CuGa$_3$Se$_5$/MZO interfaces. CuGa$_3$Se$_5$/MZO PV device with CdS (a, c, e), and Cd$^{2+}$ solution treatment (b, d, f).

The device performances were dependent on the temperature and the duration of the Cd$^{2+}$ treatment (Figure 7). Lower treatment temperature (65°C 15 min) improved the V$_{OC}$ up to 925 mV, while higher temperature (85°C 7 min) treatment resulted in higher J$_{SC}$ up to 8.3 mA/cm$^2$ (Figure 7c and Table 3). The highest open circuit voltage of 925 mV was observed for Mg composition of 10% in MZO (Figure 7a). Comparing QE of different device configurations (Figure 7b) showed that replacing CdS with Mg$_x$Zn$_{1-x}$O improved the carrier collection in the short wavelengths due to the higher bandgap of MZO. As shown previously, the conductivity of Mg$_x$Zn$_{1-x}$O could be improved by doping with Ga, which required higher substrate temperature. However, device performance was significantly reduced with V$_{OC}$ in the range of 400 to 500 mV and many devices with doped MZO were shunted (data not shown), possibly due to the higher temperature exposure.

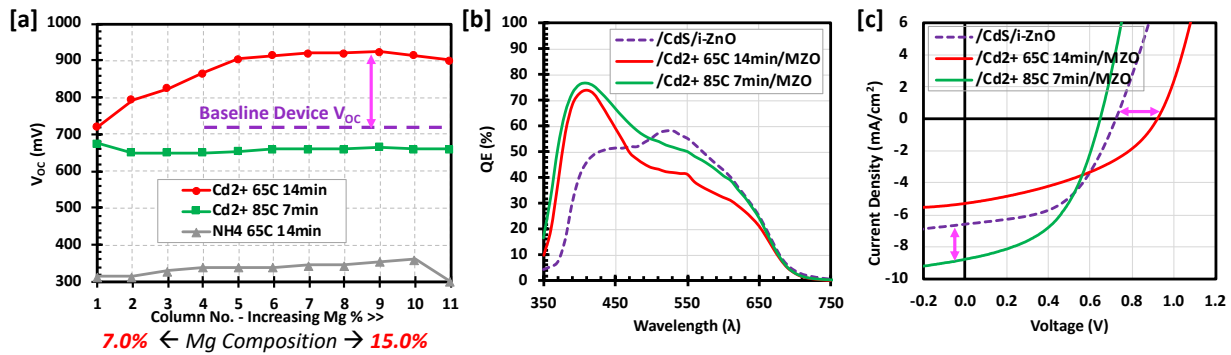

Figure 7. PV performance for CdS-free CuGa$_3$Se$_5$/Mg$_x$Zn$_{1-x}$O devices with different surface treatments, including [a] open circuit voltage, [b] external quantum efficiency, and [c] current voltage characteristics.

The photovoltaic device performance for various configurations are summarized in Table 3. Although significant open-circuit voltage improvement was observed in MZO based devices, the fill factor and photocurrent output were reduced. The CuGa$_3$Se$_5$/(Cd$^{2+}$)/MZO photovoltaic devices had superior open-circuit voltage of 925 mV compared to 730 – 755 mV for CdS based devices. Looking at the complete device structure, CuGa$_3$Se$_5$/(Cd$^{2+}$)/Mg$_x$Zn$_{1-x}$O/Al:ZnO/metal grid, the n-type counterpart for the p-type CuGa$_3$Se$_5$ absorber in this case is likely Mg$_x$Zn$_{1-x}$O/Al:ZnO bilayer. Change of absorber surface conductivity type to either intrinsic or slightly n-type could create a better p-i-n junction that improved the device V$_{OC}$, where bulk of the CuGa$_3$Se$_5$ absorber is the p-type layer, the Mg$_x$Zn$_{1-x}$O/Al:ZnO is the n-type bilayer stack, and the Cd$^{2+}$-treated CuGa$_3$Se$_5$ surface is the intrinsic (i) layer.

Table 3. Photovoltaic device performance for different configurations

| Device Configuration | V$_{OC}$ (mV) | J$_{SC}$ | FF (%) | Efficiency (%) |
|---|---|---|---|---|
| CGS/CdS/i-ZnO | 730 | 5.7 | **61.4** | **2.6** |
| CGS/CdS/MZO | 755 | 7.6 | 38.4 | 2.1 |
| CGS/Cd$^{2+}$ 65°C 14min/MZO | **925** | 4.8 | 41.9 | 1.9 |
| CGS/Cd$^{2+}$ 85°C 7min/MZO | 650 | **8.3** | 47.6 | **2.6** |

Mg$_x$Zn$_{1-x}$O contacted CuGa$_3$Se$_5$ absorber PV device data fulfilled two important requirements for top cell tandem device application: high V$_{OC}$ and improved QE in the blue region of the spectrum. Although direct water splitting tests were not performed yet due to the instability of MZO in acidic solutions, LSV testing in sacrificial redox couple showed promising outcomes. A thin protective layer deposited on top of MZO could improve stability and facilitate water splitting experiments. More experiments are in progress to integrate such protective coatings on MZO without degrading underlying absorber/contact interface, and will be reported in the future.

## 4. Conclusion:

Mg$_x$Zn$_{1-x}$O contact integration with CuGa$_3$Se$_5$ absorber has been demonstrated. MZO deposition and absorber surface treatment parameters are determined for improved photovoltaic device performance, which allowed the elimination of the CdS contact layer. Characterization of the CuGa$_3$Se$_5$ films with Cd$^{2+}$ solution surface treatment indicated that the beneficial effect of the treatment is due to removal of surface oxidation and change in surface doping by Cd substitution. The photoelectrochemical characteristics of the devices were promising for future water splitting applications: from linear sweep voltammetry in 10 mM hexaammineruthenium (III) chloride, a photocurrent onset potential near 1V vs RHE was observed. For solid state PV devices, replacing CdS/i-ZnO with MZO improved the carrier collection in the short wavelengths and resulted in open circuit voltage of 925 mV, which is promising for top-cell tandem PV device applications. The results of this research will facilitate the understanding of CuGa$_3$Se$_5$/Mg$_x$Zn$_{1-x}$O interface, and the use of CuGa$_3$Se$_5$ absorbers for both PEC and PV device applications.

**Acknowledgements**

This work was authored in part by the National Renewable Energy Laboratory (NREL), operated by Alliance for Sustainable Energy LLC, for the U.S. Department of Energy (DOE) under contract no. DE-AC36-08GO28308. Funding provided by the Office of Energy Efficiency and Renewable Energy (EERE), under Fuel Cell Technologies Office (FCTO), as a part of HydroGEN Energy Materials Network (EMN) consortium project administered by the University of Hawaii under contract no. DE-EE0006670. The authors would like to acknowledge John Perkins and

# Supplementary Information

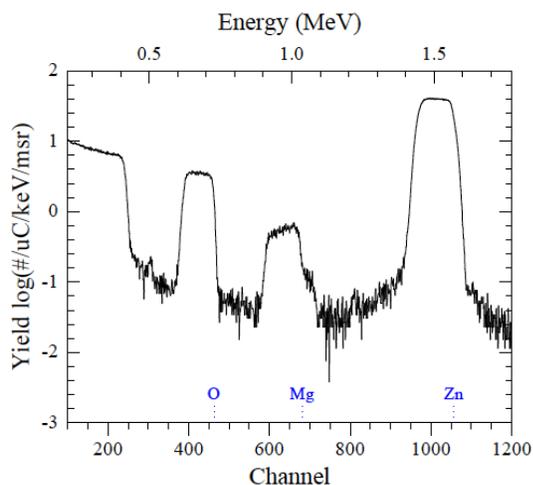

Figure S1. Rutherford Backscattering Spectroscopy (RBS) to quantify Mg composition of the MZO films.

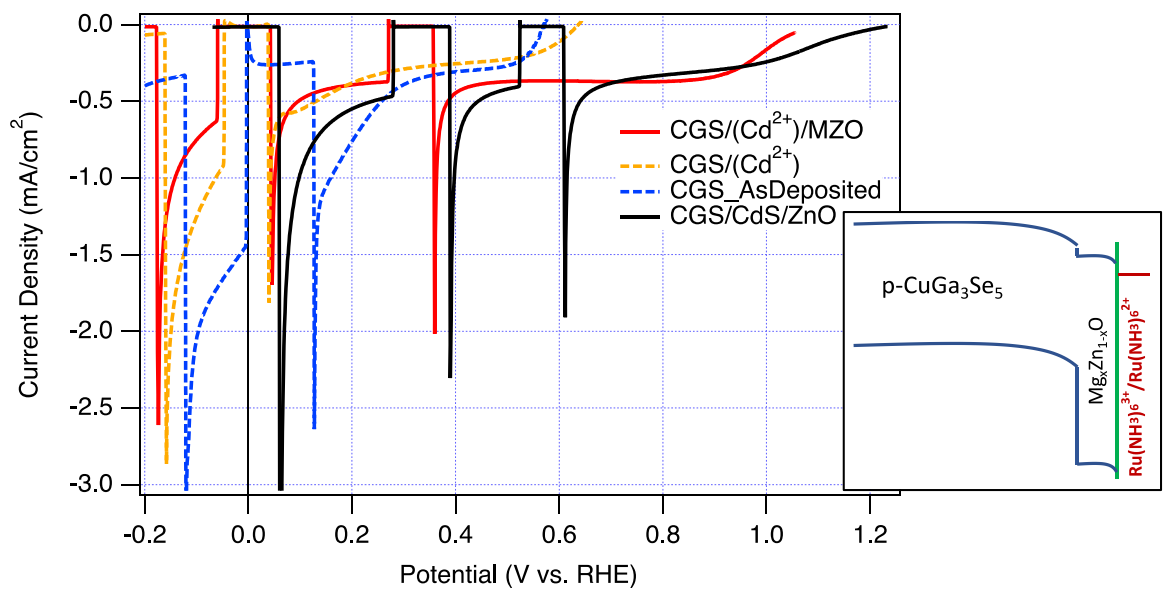

Figure S2. 3-electrode chopped light LSV measurement (in a solution containing 10 mM hexaammineruthenium (III) chloride, 0.5M KCl and pH7 buffer) data for CuGa$_3$Se$_5$ with different surface treatments and device configurations. (inset) A schematic of the equilibrium band diagram for CuGa$_3$Se$_5$/Mg$_x$Zn$_{1-x}$O photoelectrode in sacrificial redox.

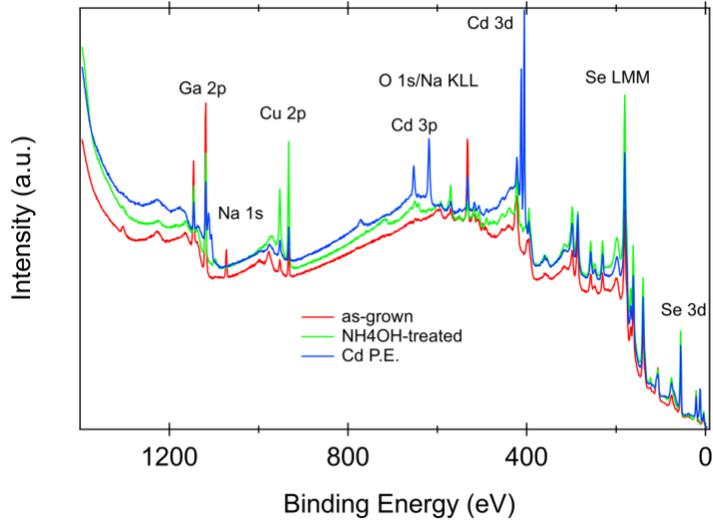

Figure S3. XPS survey spectrum of the as-deposited $CuGa_3Se_5$ film.

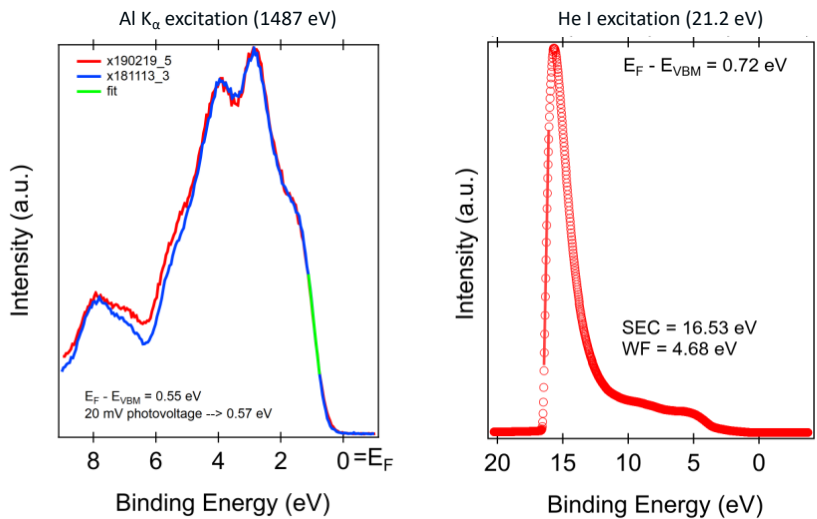

Figure S4. VBM position for as-deposited $CuGa_3Se_5$ film surface with Al $K_\alpha$ and He I excitation.

**$CuGa_3Se_5$/CdS/$Mg_xZn_{1-x}O$ devices:** A contact layer of $Mg_xZn_{1-x}O$ with CdS improved the voltage and current output of the solar cells. Both $V_{OC}$ and $J_{SC}$ of the devices were function of the Mg composition in MZO (Figure S5). At low Mg composition, the performance was equivalent to the baseline device with i-ZnO contact layer. With increasing Mg concentration, both $V_{OC}$ and $J_{SC}$ of the devices increased. Superior current output was reflected in the quantum efficiency (QE) data, where the overall QE of the devices were improved. At the longer wavelengths, carrier collection increased with increasing Mg concentration. Improved $V_{OC}$ indicated favorable conduction band offset between CdS and MZO. Improved QE implied reduction in interface carrier recombination. However, the overall device photo-conversion efficiency was reduced due to lower fill factor. The best result for $CuGa_3Se_5$/CdS/MZO devices: $V_{OC}$ = 755 mV, $J_{SC}$ = 7.6 mA/cm$^2$, FF = 38.4%, and efficiency 2.1%

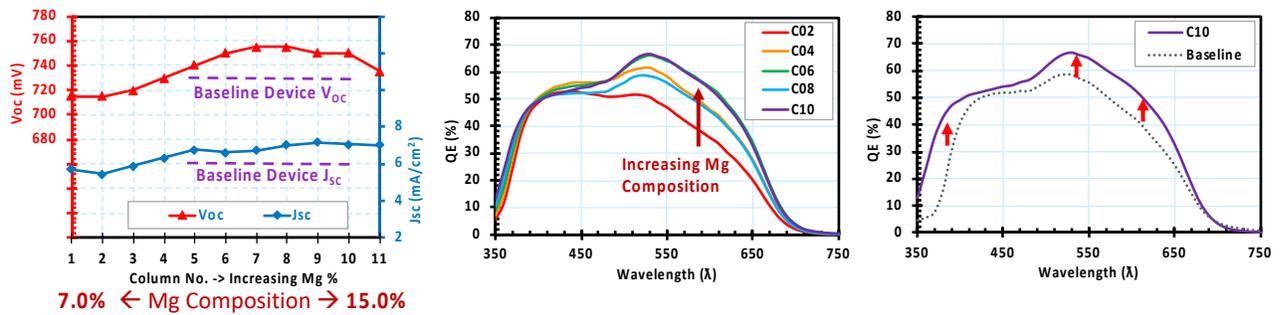

Figure S5. PV performance for $CuGa_3Se_5$/CdS/$Zn_{1-x}Mg_xO$ devices.

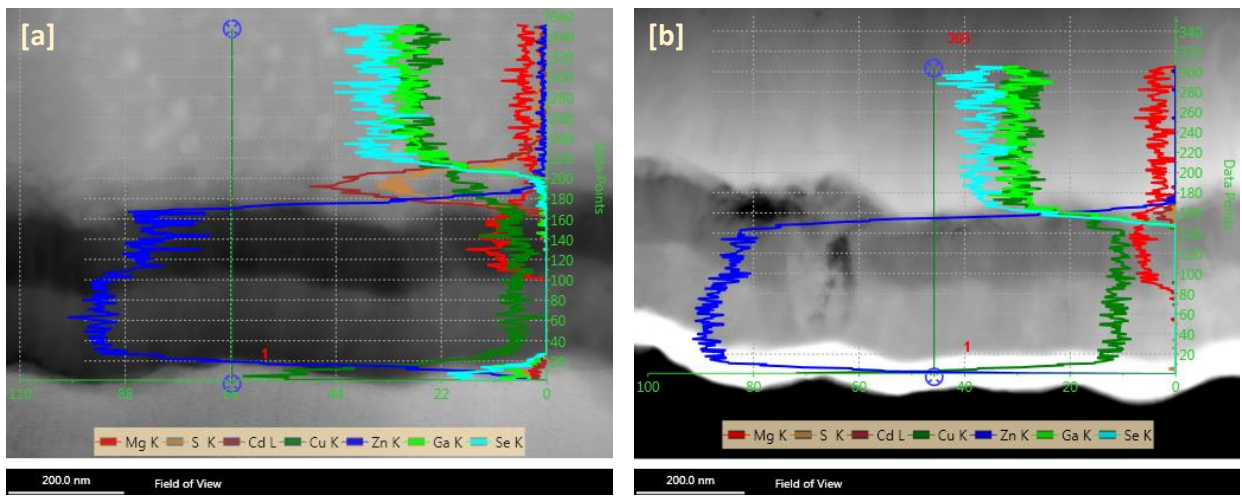

Figure S6. Elemental line profile from STEM/EDX overlaid on the STEM/HAADF image for $CuGa_3Se_5$/MZO PV device with CdS [a], and $Cd^{2+}$ solution treatment [b].